\documentclass[traditabstract]{aa}  
\usepackage{graphicx}
\usepackage{txfonts}

\begin{document}

\title{The subdwarf B star SB\,290 -- A fast rotator on the extreme horizontal branch}

\author{S. Geier \inst{1,2}
   \and U. Heber \inst{1}
   \and C. Heuser \inst{1}
   \and L. Classen \inst{1}
   \and S. J. O'Toole \inst{3} 
   \and H. Edelmann \inst{1}
   }

\offprints{S.\,Geier,\\ \email{sgeier@eso.org}}

\institute{Dr. Karl Remeis-Observatory \& ECAP, Astronomical Institute, Friedrich-Alexander University Erlangen-Nuremberg, Sternwartstr. 7, D 96049 Bamberg, Germany
\and European Southern Observatory, Karl-Schwarzschild-Str. 2, 85748 Garching, Germany  
\and Australian Astronomical Observatory, PO Box 915, North Ryde NSW 1670, Australia}

\date{Received \ Accepted}

\abstract{Hot subdwarf B stars (sdBs) are evolved core helium-burning stars with very thin hydrogen envelopes. In order to form an sdB, the progenitor has to lose almost all of its hydrogen envelope right at the tip of the red giant branch. In close binary systems, mass transfer to the companion provides the extraordinary mass loss required for their formation. However, apparently single sdBs exist as well and their formation is unclear since decades. The merger of helium white dwarfs leading to an ignition of core helium-burning or the merger of a helium core and a low mass star during the common envelope phase have been proposed. Here we report the discovery of SB\,290 as the first apparently single fast rotating sdB star located on the extreme horizontal branch indicating that those stars may form from mergers. 
}

\maketitle

\section{Introduction \label{sec:intro}}

Hot subdwarf B stars (sdBs) are core helium-burning stars with thin hydrogen envelopes situated at the extreme end of the blue horizontal branch (Heber \cite{heber09}). Horizontal branch stars are normally formed after the ignition of core helium-burning in the red-giant phase. However, the hydrogen envelopes of sdBs are extremely thin and large mass-loss just at the tip of the red-giant branch is necessary to form them. 

Ejection of a common envelope can be held responsible for the strong mass loss to form sdB stars in close binaries, which make up for half of the sdB population. While stable Roche lobe overflow explains the formation of sdB stars with main sequence companions in wider orbits, the existence of single sdB stars still remains a riddle. The most popular scenario invokes mergers of helium-core white dwarfs (Webbink \cite{webbink84}; Iben \& Tutukov \cite{ibentutukov84}). Politano et al. (\cite{politano08}) proposed the merger of a red giant and a low-mass main sequence star during the common envelope phase as another possible formation channel. 

An important constraint for evolutionary scenarios is the distribution of the rotational properties of sdB stars. Geier et al. (\cite{geier10}) have studied a sample of close binary sdB stars and found that short-period systems (orbital periods $P<1.2\,{\rm d}$) show signifi\-cant rotation while the longer period systems are slow rotators. Hence tidal interaction with the companion is important for the $v_{\rm rot}\sin{i}$-distribution of binary sdBs and has spun up the sdB stars in the closest systems.  

In order to avoid tidal interaction effects Geier \& Heber (\cite{geier12}) extended their study to 105 apparently single and wide binary sdB stars and found all of them to be slowly rotating at $v_{\rm rot}\sin{i}<10\,{\rm km\,s^{-1}}$. This result challenges merger scenarios as it would require rapid loss of angular momentum after the merger event. 

The only exception known is EC\,22081$-$1916, which rotates at $v_{\rm rot}\sin{i}=163\,{\rm km\,s^{-1}}$. There is no hint for any radial veloci\-ty variations, excluding a close companion and therefore tidal interactions. However, this star differs from typical sdB stars on the extreme horizontal branch (EHB) as its surface gravity is an order of magnitude lower. It must be an evolved object and may have formed quite differently from EHB stars (Geier et al. \cite{geier11a}).

In the course of our studies of sdB stars at high spectral resolution (Geier et al. \cite{geier10}; Geier et al. \cite{geier11a}; Geier \& Heber \cite{geier12}; Geier \cite{geier13}) the sdB star SB\,290 caught our attention because of its unusual line shapes. 

SB\,290 (CD$-$38\,222) was discovered as a B-type star in the objective prism survey of the south Galactic pole region by Slettebak \& Brundage (\cite{slettebak71}). Graham \& Slettebak (\cite{graham73}) classified the spectrum of SB\,290 as sdB. Its atmospheric parameters were derived from the spectral energy distribution including IUE UV
spectrophotometric data and optical spectra (Heber et al. \cite{heber84}). The results ($T_{\rm eff}=28200\pm1300\,{\rm K}$, $\log{g}=5.5\pm0.2$ and helium abundance $\log{y}=-2.4$) place SB\,290 on the EHB, as expected for a prototypical, that is core-helium burning, sdB star. 

First indication that SB 290 is not as typical as it seemed at first glance came from high-resolution optical spectra revealing a helium isotopic anomaly. Isotopic line shifts were found indicating that the $^{4}$He isotope in the atmosphere is almost completely replaced by $^{3}$He due to gravitational settling (Heber \cite{heber87}).
Eventually, SB\,290 became the prototype of a small class of sdBs showing this anomaly (Edelmann et al. \cite{edelmann01}; Geier et al. \cite{geier12c}).

K\"ugler (\cite{kuegler91}) noted significant broadening of the Balmer line cores of SB\,290. O'Toole (\cite{otoole04}) analysed high-resolution UV-spectra of SB\,290 obtained with the IUE satellite and also found the metal lines to be significantly broadened compared to the other sdBs in his sample. 

\begin{figure}[t!]
\begin{center}
	\resizebox{8cm}{!}{\includegraphics{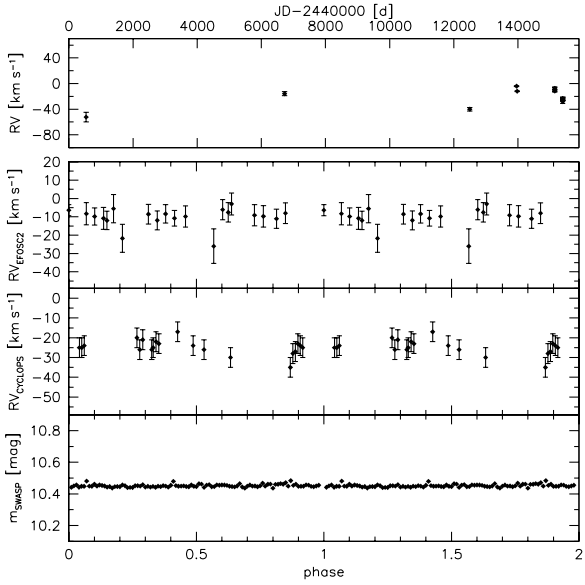}}
\end{center}
\caption{Time resolved photometry (SWASP, lower panel) and radial velocities of the CYCLOPS and EFOSC2 spectra (middle panels) plotted against phase. The data has been folded to the upper limit for the orbital period of a putative close binary with synchronised rotation of the sdB  ($P\simeq0.22\,{\rm d}$, see Geier et al. \cite{geier11a}). Two complete phases are plotted for better visualisation. Since no significant variations can be seen in the data, SB\,290 is not in a close binary system. The nightly average values of the radial velocities measured from all spectra is plotted against the Julian Date (upper panel). The oldest RV point dates back to 1969 and is taken from Graham \& Slettebak (\cite{graham73}).}
\label{nondetect}
\end{figure}

\section{Possible line broadening mechanisms}

Following the discussion in Geier et al. (\cite{geier11a}), there are different mechanisms that may cause such a broadening. High magnetic fields that may cause a Zeeman-splitting of the lines as observed in magnetic white dwarfs can be excluded. Most recently, Mathys et al. (\cite{mathys12}) measured the magnetic field of SB\,290 using the FORS2 instrument at ESO-VLT in spectropolarimetric mode. Although they report a marginal detection of a few hundred kG, this field strength is much too small to cause the observed line broadening.

A plausible explanation would be a close companion to SB\,290. Hot subdwarfs with high measured projected rotational velocities ($v_{\rm rot}\sin{i}>100\,{\rm km\,s^{-1}}$) mostly reside in very close binary systems with orbital periods of $\simeq0.1\,{\rm d}$. Those sdBs were spun up by the tidal influence of their close companions and their rotation became synchronised to their orbital motion (e.g. Geier et al. \cite{geier07,geier10}).

The colours of SB\,290 ($J-K_{S}\simeq0.06$, 2MASS, Skrutskie et al. \cite{skrutskie06}) do not show any signs of a cool companion (Stark \& Wade \cite{stark03}). A putative close companion, which is also not visi\-ble in the optical spectra, must therefore be either a low-mass main sequence star (with spectral type later than about M2), a compact object like a white dwarf or a substellar object. While a close white dwarf or M-star companion would cause RV variations of the order of tens or hundreds of ${\rm km\,s^{-1}}$, any cool companion with a size similar to the sdB primary, no matter if it would be stellar or substellar, would be detectable in the light curve. Close sdB+dM or BD systems are often eclipsing, but also show sinusoidal variations due to light orginating from the irradiated surface of the cool companion (e.g. Geier et al. \cite{geier11c}).

Furthermore, it can be premature to conclude that line broadening is due to rotation, since many sdB stars show multi-periodic pulsations with short periods ($2-10\,{\rm min}$), which can result in line broadening as high as $34\,{\rm km\,s^{-1}}$ in time-integrated spectra and extreme cases (Kuassivi et al. \cite{kuassivi05}).

Therefore we extracted photometric measurements from the SuperWASP archive and searched for short-period light variations. If the broadening were indeed due to rotation we would expect that SB\,290 has been spun up by tidal interaction with a close companion. Assuming spin-orbit synchronisation, we would expect an orbital period of $\simeq0.2\,{\rm d}$. Therefore, we also obtained time-resolved spectroscopy to verify the existence of a nearby companion.

\section{Observations, light curve and radial velocity curve \label{sec:obs}}

\subsection{Photometry}

Due to its brightness, SB\,290 has been monitored by planetary transit surveys. An excellent white light curve taken from May 2006 to December 2007 was downloaded from the SuperWASP Public archive\footnote{http://www.wasp.le.ac.uk/public/lc/index.php} (Pollacco et al. \cite{pollacco06}). The light curve consists of no less than $10192$ single measurements. 

In order to search for periodic variations in the SuperWASP light curve, we performed a Fourier-analysis using the FAMIAS routine developed by Zima (\cite{zima08}). No significant light variations have been detected (see Fig.~\ref{nondetect}, lower panel). A cool low-mass stellar or substellar companion in close orbit can therefore be excluded. Although the uneven sampling of the light curve does not allow to perform a proper search for pulsations, strong variations causing significant line broadening can be excluded as well.

\subsection{Time resolved spectroscopy}

Time-resolved medium-resolution spectroscopy ($R\simeq4000,\lambda=3500-5100\,{\rm \AA}$) was obtained in the course of the MUCHFUSS project (Geier et al. \cite{geier11b}). One dataset consisting of 19 spectra was taken with the ISIS spectrograph mounted at the WHT in August 2009. Reduction was done with standard IRAF procedures. Another set of 20 single spectra ($R\simeq2200,\lambda=4450-5110\,{\rm \AA}$) was obtained with the EFOSC2 spectrograph mounted at the ESO-NTT in November 2009. Reduction was done with standard MIDAS procedures. Finally, 34 high resolution spectra ($R\simeq70000,\lambda=3900-5270\,{\rm \AA}$) were obtained with the UCLES spectrograph equipped with the CYCLOPS fibre feed\footnote{For some details on this instrument see http://www.phys.unsw.edu.au/$\sim$cgt/CYCLOPS/CYCLOPS\_Classic.html} and mounted at the AAT in July 2010.

Radial velocities were measured by fitting a set of mathematical functions (polynomial, Lorentzian and Gaussian) to the Balmer and suitable helium lines of the spectra using the FITSB2 routine (Napiwotzki et al. \cite{napiwotzki04}). No significant RV variations were measured within the EFOSC2 and CYCLOPS datasets.\footnote{The ISIS spectra showed a linear trend in RV most likely caused by flexure of the instrument, because the target was observed at very high zenith distance.} The radial velocity of the CYCLOPS spectra is constant at $-25\pm4\,{\rm km\,s^{-1}}$, the one of the EFOSC2 spectra at $-10\pm5\,{\rm km\,s^{-1}}$, where the standard deviations of the respective single measurements are adopted as uncertainties (see Fig.~\ref{nondetect}, middle panels). Since no RV variations have been measured on timescales of a few hours, we can exclude any close stellar companion.

\subsection{Long-term RV variations}

While there is only a small discrepancy between the RVs measured from the CYCLOPS and EFOSC2 datasets, which might have originated from zero-point shifts between the two instruments, Graham \& Slettebak (\cite{graham73}) measured a significantly different RV of $-52.3\pm7.5\,{\rm km\,s^{-1}}$. In order to investigate the possibility of an unseen companion in a wide orbit, we measured the RVs from spectra taken within the last 25 years (see Table~\ref{RV}). 

A high resolution spectrum was taken back in 1986 with the CASPEC spectrograph ($R=30\,000,\lambda=3800-5000\,{\rm \AA}$) mounted at the ESO 3.6m telescope at La Silla. Two high resolution spectra were taken in 2002 with the FEROS spectrograph ($R=48\,000,\lambda=3750-9200\,{\rm \AA}$) mounted at the ESO 1.52m telescope and another two spectra were taken with FEROS mounted at the ESO/MPG\,2.2\,m telescope at La Silla in 2006. The spectra have been reduced with the MIDAS package. 

The spectra taken with CASPEC and FEROS indeed show a maximum RV-shift of $\simeq25\,{\rm km\,s^{-1}}$ within $16\,{\rm yr}$. SB\,290 might therefore have an unseen companion in a wide orbit (see Fig.~\ref{nondetect}, upper panel), which is, however, not likely to be responsible for the observed line broadening of the sdB.

\begin{figure}[t!]
\begin{center}
	\resizebox{7cm}{!}{\includegraphics{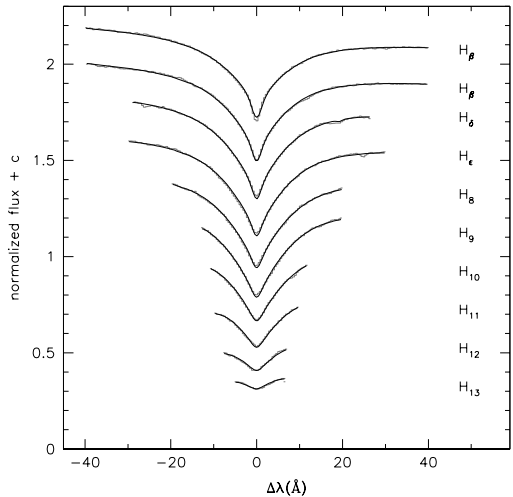}}
\end{center}
\caption{Fit of synthetic LTE models to the hydrogen Balmer lines of the medium resolution ISIS spectrum.}
\label{fitH}
\end{figure}

\begin{figure}[t!]
\begin{center}
	\resizebox{7cm}{!}{\includegraphics{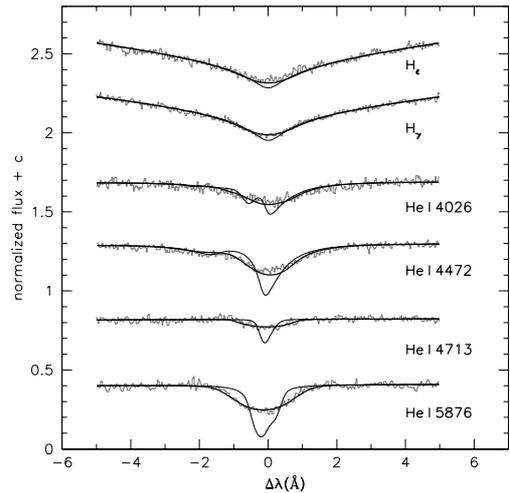}}
\end{center}
\caption{Fit of synthetic LTE models to selected hydrogen Balmer and helium lines of the high-resolution FEROS spectrum. The thin solid line marks models without rotational broadening, the thick solid line the best fitting model spectrum with $v_{\rm rot}\sin{i}=58\,{\rm km\,s^{-1}}$. The significant rotational broadening of the lines is obvious.}
\label{fitHe}
\end{figure}

\section{Atmospheric parameters and rotational broadening \label{atmo}}

Atmospheric parameters have been determined by fitting a grid of synthetic spectra, calculated from line-blanketed, solar-metalicity LTE model atmospheres (Heber et al. \cite{heber00}), to the hydrogen Balmer (H$_{\rm \beta}$-H$_{\rm 13}$) and helium lines (He\,{\sc i}\,4026, 4472, 4713, 4922, 5016\,${\rm \AA}$) of the coadded ISIS spectrum using the SPAS routine developed by H. Hirsch (e.g. Geier et al. \cite{geier11b}). Statistical errors are determined with a bootstrapping algorithm. 

The resulting effective temperature $T_{\rm eff}=26\,300\pm100\,{\rm K}$, surface gravity $\log{g}=5.31\pm0.01$ and helium abundance $\log{y}=-2.52\pm0.08$ are characteristic for sdB stars. 

In order to measure the projected rotational velocity, we used the coadded high-resolution FEROS spectrum, which is very well suited for this purpose (see Geier et al. \cite{geier10}; Geier \& Heber \cite{geier12}). The atmospheric parameters have been fixed to the values derived from the ISIS spectrum and only the $v_{\rm rot}\sin{i}$ has been fitted. As can be seen in Fig.~\ref{fitHe}, the helium lines are significantly broadened. In order to fit those lines as well as the Balmer line cores, a broadening of $58\pm1\,{\rm km\,s^{-1}}$ is necessary. After excluding all alternative scenarios we conclude that this broadening is caused by rotation.

In addition to Balmer and helium lines, the spectrum shows several metal lines. The C\,{\sc ii} lines at $4267\,{\rm \AA}$ as well as the Si\,{\sc iii} lines at $4552$, $4567$, $4574$ and $5739\,{\rm \AA}$ are strong enough to measure both the abundances and the rotational broadening in the way described in Geier (\cite{geier13}). The $v_{\rm rot}\sin{i}=48\pm2\,{\rm km\,s^{-1}}$ measured from those lines is somewhat smaller than the one measured from the Balmer and helium lines. 

Geier (\cite{geier13}) reported the discovery of C\,{\sc ii} and Si\,{\sc iii} lines with peculiar shapes in sdB stars, which also show the $^{3}$He anomaly. The observed lines appeared narrower than the synthetic line profiles, even without any rotational broadening. Vertical stratification in the atmosphere may be responsible. Since SB\,290 is the prototype for the $^{3}$He-sdBs, it might also be affected. However, due to the high line broadening, the effect might be less obvious and only lead to an underestimate of $v_{\rm rot}\sin{i}$. In this special case, we therefore regard the rotational velocity derived from the Balmer and helium lines to be more reliable. The abundances of carbon ($\log{\epsilon}_{\rm C}=7.10\pm0.30$) and silicon ($\log{\epsilon}_{\rm Si}=7.28\pm0.15$) are typical for sdBs in this temperature range (Geier \cite{geier13}).

Adopting typical systematic uncertainties of $\pm500\,{\rm K}$ in $T_{\rm eff}$ and $\pm0.05\,{\rm dex}$ in $\log{g}$, these results are also roughly consistent with the previous determination by Heber et al. (\cite{heber84}), although the effective temperature determined here is somewhat lower.

\begin{figure}[t!]
\begin{center}
	\resizebox{7.5cm}{!}{\includegraphics{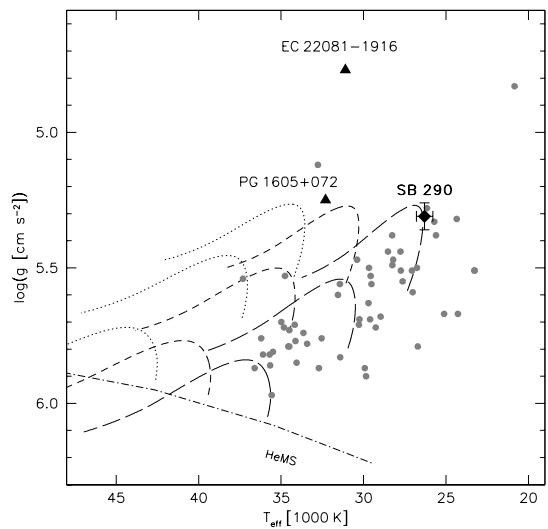}}
\end{center}
\caption{$T_{\rm eff}-\log{g}$ diagram. The grey circles mark sdBs from the SPY project (Lisker et al. \cite{lisker05}). The helium main sequence (HeMS) is plotted and evolutionary tracks have been calculated by Han et al. (\cite{han02}) for EHB masses of $0.7\,M_{\rm \odot}$ (dotted lines), $0.6\,M_{\rm \odot}$ (short-dashed lines) and $0.5\,M_{\rm \odot}$ (long-dashed lines). The tracks to the lower left mark models without hydrogen envelope, the tracks in the middle models with $0.001\,M_{\rm \odot}$ envelope mass and the tracks to the upper right models with $0.005\,M_{\rm \odot}$ envelope mass.}
\label{tefflogg}
\end{figure}

\section{Discussion}

The fact that SB\,290 is a fast rotating sdB star might be an indication for a merger origin, although empirical evidence grows that the bulk of the known single sdB stars was not formed by merger events (e.g. Fontaine et al. \cite{fontaine12}; Geier \& Heber \cite{geier12}). The fast rotator EC\,22081$-$1916 on the other hand, might be the outcome of a rare common envelope merger proposed by Politano et al. (\cite{politano08}). Since the outcome of a He-WD merger should have no hydrogen envelope at all, while the atmosphere of SB\,290 is hydrogen rich, the common envelope merger might be a possible formation channel for SB\,290 as well. 

Hot subdwarf stars formed by a merger event are predicted to have a broad mass distribution with possible masses of up to $\simeq0.7\,M_{\rm \odot}$. Han et al. (\cite{han02}) calculated evolutionary tracks for EHB stars with different stellar and envelope masses. SB\,290 is situated close to the terminal-age EHB if its mass should be close to the canonical value (Fig.~\ref{tefflogg}). Looking at Fig.~\ref{tefflogg} one can see that the tracks for an EHB mass of $0.7\,M_{\rm \odot}$ hardly match the position of SB\,290 in the $T_{\rm eff}-\log{g}$-diagram. With a mass of $0.6\,M_{\rm \odot}$ SB\,290 would on the other be situated close to the zero-age EHB.

However, it might still be premature to exclude a high mass right away. An object closely related to SB\,290 might be the strong sdB pulsator PG\,1605$+$072, which is known to be a single star with substantial line broadening ($v_{\rm rot}\sin{i}=39\,{\rm km\,s^{-1}}$, Heber \cite{heber99}). Although this broadening is most likely caused by unresolved pulsations rather than substantial rotation (Langfellner et al. \cite{langfellner12}), asteroseismic analyses indicate a high mass of more than $0.7\,M_{\rm \odot}$ (van Spaandonk et al. \cite{vanspaandonk08}; van Grootel et al. \cite{vangrootel10}). These results are still under debate, but the tracks for stellar masses of $0.7\,M_{\rm \odot}$ and different masses of the hydrogen envelope are more or less consistent with the position of PG\,1605$+$072 in the $T_{\rm eff}-\log{g}$-diagram.

The confirmation of a companion in a wide orbit might complicate the situation since it would require the progenitor to have been in a hierarchical triple system before. On the other hand, if SB\,290 would have originated from such a triple system, the putative outer companion may have accelerated the merger process by orbit shrinkage through Kozai cycles (Fabrycky \& Tremaine \cite{fabrycky07}). 

\begin{acknowledgements}

S.~G. was supported by the Deutsche Forschungsgemeinschaft under grant He~1354/49-1. The authors thank L. Morales-Rueda for providing her data and Z. Han for providing his evolutionary tracks. We would also like to thank P. Butler for helping us with the reduction of the CYCLOPS data. Furthermore, we thank the referee P. Bergeron. Based on observations at the La Silla Observatory of the European 
Southern Observatory for programmes number 077.D-0515(A) and 084.D-0348(A). Based on observations at the Australian Astronomical Observatory. Based on observations with the William Herschel Telescope operated by the Isaac Newton Group at the Observatorio del Roque de los Muchachos of the Instituto de Astrofisica de Canarias on the island of La Palma, Spain.

\end{acknowledgements}

\newpage

\begin{appendix}

\begin{table*}
\caption{Radial velocities of SB\,290}
\label{RV}
\begin{center}
\begin{tabular}{lrl}
\hline
\noalign{\smallskip}
mid$-$HJD & RV [${\rm km\,s^{-1}}$] & Instrument\\
$-2\,400\,000$ & & \\
\noalign{\smallskip}
\hline
\noalign{\smallskip}
40539.50\tablefootmark{1} & $-52.3\pm7.5$ & Graham \& Slettebak \cite{graham73} \\
\noalign{\smallskip}
\hline
\noalign{\smallskip}
46724.57    & $-15.9\pm3.2$  & CASPEC \\
\noalign{\smallskip}
\hline
\noalign{\smallskip}
52495.66123 & $-40.4\pm3.4$  & FEROS \\
52495.70637 & $-39.5\pm4.5$  & \\
53953.84358 & $-4.1\pm0.6$   & \\ 
53974.88413 & $-12.0\pm0.7$  & \\
\noalign{\smallskip}
\hline
\noalign{\smallskip}
55144.62633 &   $-6.4\pm3.1$ & EFOSC2 \\
55144.66202 &   $-12.0\pm5.0$ & \\ 
55145.62589 &   $-8.5\pm5.4$ & \\ 
55145.62803 &   $-11.9\pm5.2$ & \\ 
55145.63016 &   $-9.1\pm5.8$ & \\ 
55145.63229 &   $-9.7\pm5.8$ & \\ 
55145.63443 &   $-11.0\pm5.3$ & \\ 
55145.63656 &   $-8.0\pm5.6$ & \\ 
55145.66389 &   $-8.3\pm6.1$ & \\ 
55145.66602 &   $-9.8\pm4.7$ & \\ 
55145.66815 &   $-10.8\pm6.0$ & \\ 
55145.72750 &   $-5.5\pm7.7$ & \\ 
55145.72998 &   $-21.7\pm7.6$ & \\ 
55145.73247 &   $-26.0\pm9.4$ & \\ 
55146.59354 &   $-6.1\pm5.5$ & \\ 
55146.59602 &   $-3.0\pm6.0$ & \\ 
55146.59851 &   $-8.4\pm5.0$ & \\ 
55146.62859 &   $-10.8\pm4.3$ & \\ 
55146.65136 &   $-9.8\pm5.7$ & \\ 
55146.67425 &   $-7.5\pm5.3$ & \\ 
\noalign{\smallskip}
\hline
\noalign{\smallskip}
55393.22067 &  $-26\pm5$ &  CYCLOPS\tablefootmark{2} \\
55393.25356 &  $-24\pm5$ &  \\
55393.28933 &  $-35\pm5$ &  \\
55394.17695 &  $-28\pm5$ &  \\
55394.26083 &  $-27\pm5$ &  \\
55394.26853 &  $-23\pm5$ &  \\ 
55394.27977 &  $-24\pm5$ &  \\ 
55394.28749 &  $-25\pm5$ &  \\ 
55394.33581 &  $-25\pm5$ &  \\ 
55395.22288 &  $-25\pm5$ &  \\ 
55395.23057 &  $-24\pm5$ &  \\ 
55395.23918 &  $-25\pm5$ &  \\ 
55395.24687 &  $-22\pm5$ &  \\ 
55395.32561 &  $-23\pm5$ &  \\ 
55395.33330 &  $-20\pm5$ &  \\ 
55395.34099 &  $-26\pm5$ &  \\ 
55396.16408 &  $-21\pm5$ &  \\ 
55396.17177 &  $-17\pm5$ &  \\ 
55396.18130 &  $-26\pm5$ &  \\ 
55396.21822 &  $-30\pm5$ &  \\ 
\noalign{\smallskip}
\hline
\end{tabular}
\end{center}
\tablefoottext{1}{The observing time could only be reconstructed to within this night.}
\tablefoottext{2}{Due to the highly variable S/N-ratio of the single spectra, consecutive spectra have been coadded until the quality was sufficient to measure the RV to within about the given uncertainty. In this way, 20 RV epochs have been obtained.}
\end{table*}

\end{appendix}


\begin{thebibliography}{}

\bibitem[2001]{edelmann01} Edelmann, H., Heber, U., \& Napiwotzki, R. 2001, AN, 322, 401
\bibitem[2007]{fabrycky07} Fabrycky, D., \& Tremaine, S. 2007, ApJ, 669, 1298
\bibitem[2012]{fontaine12} Fontaine, G., Brassard, P., Charpinet, S., et al. 2012, A\&A, 539, 12
\bibitem[2013]{geier13} Geier, S. 2013, A\&A, 549, 110
\bibitem[2012]{geier12} Geier, S., \& Heber, U. 2012, A\&A, 543, 149
\bibitem[2007]{geier07} Geier, S., Nesslinger, S., Heber, U., Przybilla, N., Napiwotzki, R., \& Kudritzki, R.-P. 2007, A\&A, 464, 299 
\bibitem[2010]{geier10} Geier, S., Heber, U., Podsiadlowski, Ph., et al. 2010, A\&A, 519, 25
\bibitem[2011a]{geier11a} Geier, S., Classen, L., \& Heber, U. 2011a, ApJ, 733, L13
\bibitem[2011b]{geier11b} Geier, S., Hirsch, H., Tillich, A., et al. 2011b, A\&A, 530, 28
\bibitem[2011c]{geier11c} Geier, S., Schaffenroth, V., Drechsel, H., et al. 2011c, ApJ, 731, L22
\bibitem[2012]{geier12c} Geier, S., Heber, U., Edelmann, H., et al. 2012, ASP Conf. Ser., 452, 57
\bibitem[1973]{graham73} Graham, J. A., \& Slettebak, A. 1973, AJ, 78, 295
\bibitem[2002]{han02} Han, Z., Podsiadlowski, Ph., Maxted, P. F. L., Marsh, T. R., \& Ivanova, N. 2002, MNRAS, 336, 449
\bibitem[2003]{han03} Han, Z., Podsiadlowski, Ph., Maxted, P. F. L., \& Marsh, T. R. 2003, MNRAS, 341, 669
\bibitem[1987]{heber87} Heber, U. 1987, MitAG, 70, 79
\bibitem[2009]{heber09} Heber, U. 2009, ARA\&A, 47, 211
\bibitem[1984]{heber84} Heber, U., Hunger, K., Jonas, G., \& Kudritzki, R. P. 1984, A\&A, 130, 119
\bibitem[1999]{heber99} Heber, U., Reid, I. N., \& Werner, K. 1999, A\&A, 348, 25
\bibitem[2000]{heber00} Heber, U., Reid, I. N., \& Werner, K.\ 2000, \aap, 363, 198
\bibitem[2003]{heber03} Heber, U., Edelmann, H., Lisker, T., \& Napiwotzki, R. 2003, A\&A, 411, 477
\bibitem[1984]{ibentutukov84} Iben, I., Jr., \& Tutukov, A.~V. 1984, \apjs, 54, 335  
\bibitem[1998]{kilkenny98} Kilkenny, D., van Wyk, F., Roberts, G., Marang, F., \& Cooper, D. 1998, MNRAS, 294, 93
\bibitem[2005]{kuassivi05} Kuassivi, Bonanno, A., \& Ferlet, R. 2005, A\&A, 442, 1015 
\bibitem[1991]{kuegler91} K\"ugler, L. 1991, Diploma thesis, Christian Albrechts Universit\"at Kiel
\bibitem[2012]{langfellner12} Langfellner, J., Schuh, S., et al. 2012, ASP Conf. Ser., 452, 203
\bibitem[2005]{lisker05} Lisker, T., Heber, U., Napiwotzki, R., et al. 2005, A\&A, 430, 223
\bibitem[2012]{mathys12} Mathys, G., Hubrig, S., Mason, E., et al. 2012, AN, 333, 30
\bibitem[2004]{napiwotzki04} Napiwotzki, R., Yungelson, L., Nelemans, G. et al. 2004, ASP Conf. Ser., 318, 402
\bibitem[2004]{otoole04} O'Toole, S. J. 2004, A\&A, 423, 25
\bibitem[2008]{politano08} Politano, M., Taam, R. E., van der Sluys, M., \& Willems, B. 2008, ApJL, 687, 99
\bibitem[2006]{pollacco06} Pollacco, D. L., Skillen, I., Collier Cameron, A., et al. 2006, PASP, 118, 1407
\bibitem[2006]{skrutskie06} Skrutskie, M. F., et al. 2006, AJ, 131, 1163
\bibitem[1971]{slettebak71} Slettebak, A., \& Brundage, R. K. 1971, AJ, 76, 338
\bibitem[2003]{stark03} Stark, M. A., \& Wade, R. 2003, AJ, 126, 1455
\bibitem[2010]{vangrootel10} van Grootel, V., Charpinet, S., Fontaine, G., \& Brassard, P. 2010, Ap\&SS, 329, 217
\bibitem[2008]{vanspaandonk08} van Spaandonk, L., Fontaine, G., Brassard, P., \& Aerts, C. 2008, ASP Conf. Ser., 392, 387
\bibitem[1984]{webbink84} Webbink, R.~F.\ 1984, \apj, 277, 355 
\bibitem[2008]{zima08} Zima, W. 2008, CoAst, 157, 387

\end{thebibliography}
\end{document}